\documentclass[pra,aps,superscriptaddress,amsmath,amssymb,twocolumn,reprint]{revtex4-2}
\usepackage{CJKutf8}
\usepackage{graphicx}
\usepackage{subfigure}
\usepackage{dcolumn}
\usepackage{bm,braket}
\usepackage{mathptmx}
\usepackage{amsmath}
\usepackage{amsfonts}
\usepackage{amssymb}
\usepackage{mathrsfs}
\usepackage{float}
\usepackage{verbatim}
\usepackage{xcolor}
\usepackage[colorlinks=true,linkcolor=blue,anchorcolor=red,citecolor=blue, urlcolor=blue]{hyperref}
\usepackage{ulem}
\usepackage{pifont}

\makeatletter
\renewcommand*\env@matrix[1][\arraystretch]{%
	\edef\arraystretch{#1}%
	\hskip -\arraycolsep
	\let\@ifnextchar\new@ifnextchar
	\array{*\c@MaxMatrixCols c}}
\makeatother

\def\be{\begin{equation}}
	\def\ee{\end{equation}}
\def\ba{\begin{eqnarray}}
	\def\ea{\end{eqnarray}}

\newlength{\seplinewidth}
\newlength{\seplinesep}
\colorlet{sepline}{orange}

\begin{document}
	\begin{CJK*}{UTF8}{gbsn}
		
		\title{Time bound of atomic adiabatic evolution in the accelerated optical lattice}
		\author{Guoling Yin}
		\affiliation{State Key Laboratory of Quantum Optics and Quantum Optics Devices, Institute of Opto-Electronics, Shanxi University, Taiyuan 030006, China}
		\affiliation{State Key Laboratory of Advanced Optical Communication Systems and Networks, Department of Electronics, Peking University, Beijing 100871, China}
		
		\author{Lingchii Kong}
		\affiliation{International Center for Quantum Materials, School of Physics, Peking University, Beijing 100871, China}		
		
		\author{Zhongcheng Yu}
		\affiliation{State Key Laboratory of Advanced Optical Communication Systems and Networks, Department of Electronics, Peking University, Beijing 100871, China}
		
		\author{Jinyuan Tian}
		\affiliation{State Key Laboratory of Advanced Optical Communication Systems and Networks, Department of Electronics, Peking University, Beijing 100871, China}	

	   \author{Xuzong Chen}
		\affiliation{State Key Laboratory of Advanced Optical Communication Systems and Networks, Department of Electronics, Peking University, Beijing 100871, China}

		\author{Xiaoji Zhou}
		\email{xjzhou@pku.edu.cn}
		\affiliation{State Key Laboratory of Quantum Optics and Quantum Optics Devices, Institute of Opto-Electronics, Shanxi University, Taiyuan 030006, China}
		\affiliation{State Key Laboratory of Advanced Optical Communication Systems and Networks, Department of Electronics, Peking University, Beijing 100871, China}
		\date{\today}

		\begin{abstract} 
			The accelerated optical lattice has emerged as a valuable technique for the investigation of quantum transport physics and has found widespread application in quantum sensing, including atomic gravimeters and atomic gyroscopes. In our study, we focus on the adiabatic evolution of ultra-cold atoms within an accelerated optical lattice. Specifically, we derive a time bound that delimits the duration of atomic adiabatic evolution in the oscillating system under consideration. To experimentally substantiate the theoretical predictions, precise measurements to instantaneous band populations were conducted within a one-dimensional accelerated optical lattice, encompassing systematic variations in both lattice's depths and accelerations. The obtained experimental results demonstrate a quantitatively consistent correspondence with the anticipated theoretical expressions. Afterwards, the atomic velocity distributions are also measured to compare with the time bound. This research offers a quantitative framework for the selection of parameters that ensure atom trapped throughout the acceleration process. Moreover, it contributes an experimental criterion by which to assess the adequacy of adiabatic conditions in an oscillating system, thereby augmenting the current understanding of these systems from a theoretical perspective.
		\end{abstract}
		
		\maketitle
	\end{CJK*}	
	\section{INTRODUCTION}

Bloch electrons that are accelerated by an external static electric field have been long discussed in solid state physics since Bloch and Zener's formula in the 1930s \cite{Landau1932, Zener1932, Stuckelberg1932, Majorana1932}. Many striking phenomena are predicted due to the interplay between the quantum effects and the translation symmetry \cite{Gluck2002, Wu2001, Witthaut2005, Breid2006, Vitanov1999, Yue2010, Niu1999}. However, observing these in experiments is very challenging owing to the poor control of the lattice defects, impurities, interactions between particles and etc. in solid state materials\cite{Leo1992}.
Afterwards, due to the rise of Bose-Einstein condensate (BEC) and laser techniques, the optical lattice became an ideal system to simulate the Bloch electrons using neutral atoms with nearly all parameters being controlled \cite{Parker2013}. The 1D static optical lattice is created by two counter-propagating lasers with identical frequencies. And, the accelerated lattice can be realized by properly scanning the phase difference between the lattice beams \cite{Cadoret2008,Struck2011}.

Nowadays, the accelerated optical lattice can be used to resemble Wannier-Stark Hamiltonian in order to study Bloch oscillation, Landau-Zener transition, Wannier's ladders and etc.\cite{BenDahan1996, Peik1997, Wilkinson1996, Anderson1998, Morsch2001, Morsch2006, Zhou2018, Wilkinson1997, Grynberga2001, Zenesini2009, Tayebirad2010, Haller2010,Fujiwara2019,Yu2023, Wilkinson1997, Bharucha1997, Niu1998, Choi1999, Hartmann2004}.
Besides, the accelerated optical lattice can also be used in gravimeters, where the atoms can be accelerated against gravity in order to improve sensitivity\cite{Cadoret2008,Andia2013}. During this process, many atoms escape from the lattice and only a few are left for the following interference if the acceleration is large or the well depth is small. It can be caused by quantum tunneling and the number of escaped atoms is typically an exponential function of time. The number of left atoms is usually balanced against the time of acceleration based on researchers' experience. 

Here, we derive a time bound of atomic adiabatic evolution in an accelerated optical lattice and observe it in the experiment. Within this time bound, the atoms will be trapped in the initial band and cannot escape from the lattice. The time bound is measured with different lattice depths and lattice accelerations after using our proposed shortcut method to load atoms into the S-band of the optical lattice, and the experimental results agree well with the theoretical predictions.
Meanwhile, this time bound is derived as a part of adiabatic conditions in accelerated optical lattice which possesses an oscillating Hamiltonian. The general sufficient adiabatic conditions of such Hamiltonian are still implicit in theory. Thus, the experimental evidence of this time bound can contribute to a judge for various adiabatic conditions from different theories.

This paper is organized as follows. Section \ref{sec:2} formulates the atoms in an accelerated optical lattice. In Section \ref{sec:2A}, a time bound of adiabatic evolution is derived, where the Hamiltonian is simplified by a two-state approximation. In Section \ref{sec:2B}, the time bound is explained together with the known decay mechanisms in this system. Section \ref{sec:3} illustrates the experimental procedures of measuring the number of atoms in the instantaneous band. Section \ref{sec:4} shows the experimental results. In Section \ref{sec:4A}, the time-resolved distribution of atoms in the instantaneous S-band $P_S(t)$ with different parameters is shown  (the atoms are initially prepared in S-band). In Section \ref{sec:4B}, the times when $P_S(t)$ decreases are found and compared with the expression of time bound.  Then we give a discussion on the time bound and the atomic velocity in \ref{sec}. The conclusions are summed in \ref{sec:7}.

\section{The time bound of adiabatic evolution}\label{sec:2}
\subsection{Theoretical Model}\label{sec:2A}
The non-interacting atoms in a constantly accelerated lattice can be described by a single particle Hamiltonian, i.e.,
\begin{equation}\label{1}
	\hat{H}(t) = \frac{\hat{P}^2}{2m} + V_0\sin^2\left[k_L \left(x-\frac{at^2}{2}\right)\right]
\end{equation}
where $ \hat{P} $ is the momentum operator, $V_0$ is the well depth, $m$ is the mass of atom, $a$ is the acceleration of optical lattice, $k_L = \pi/d$ is the wave number of lasers and $d$ is the lattice constant. The atom evolves with the time-dependent Schr\"{o}dinger equation, i.e., $i\hbar \partial_t|\Psi(t)\rangle = \hat{H}(t)|\Psi(t)\rangle$ where $\Psi(t)$ is the atom's wave function.
	
Suppose the initial state is the ground state. With a small acceleration, the atom will move adiabatically, i.e., the atom will stay in the instantaneous ground state. To give a better illustration, we derive the adiabatic conditions. The wave function is expanded by momentum basis, i.e., $|\Psi(t)\rangle = \sum_{p}c_{p}(t)|p\rangle=\sum_{n,k}c_{n,k}(t)|n,k\rangle$ where the momentum $p$ is split by a reciprocal vector index $n$ and a quasi-momentum $k$. Then the entries of Hamiltonian \eqref{1} become
\begin{equation}\label{2}
	H_{n,n;k,k'} = 4n^2E_r\delta_{k,k'},\quad H_{n,n\pm 1;k,k'} = \frac{V_0}{4}e^{\pm ik_Lat^2}\delta_{k,k'}
\end{equation}
where $E_r = \hbar^2k_L^2/(2m)$ is the recoil energy. The quasi-momentum $k$ which is preserved along dynamics is set to zero for simplicity. 
If the atoms perform a Bloch oscillation, there will be a momenta transfer from $0$ to $2\hbar k_L$ within a Bloch period defined by
\begin{equation}\label{3}
	T_\mathrm{B}\equiv \frac{2\hbar k_L}{ma}.
\end{equation}
These two momenta states outperform the others during $t \lessapprox T_\mathrm{B}$. Thus, it is legal to simplify the Hamiltonian Eq.\eqref{2} using the two-state approximation.
The simplified Hamiltonian becomes
\begin{equation}\label{4}
	h=\begin{pmatrix}
		0 & {V_0}e^{ik_Lat^2}/{4}\\
		{V_0}e^{-ik_Lat^2}/{4} & 4~E_r
	\end{pmatrix}.
\end{equation}
It is similar to the Schwinger's Hamiltonian but with a quadratic time dependent form. Following the results in \cite{Comparat2009}, the adiabatic condition for this oscillating Hamiltonian can be obtained by solving inequalities
\begin{subequations}
	\label{5}
\begin{align}
	\left\|\Delta^{-1}\right\| \left\|\Omega\right\|&\ll 1 \label{5a} \\
	\int_{0}^{t}dt' \left(\left\| \Omega\right\|\left\| \frac{d\Delta^{-1}}{dt'}\right\| + \left\| \Delta^{-1}\right\|\left\| \frac{d\Omega}{dt'}\right\|\right)&\ll1 \label{5b}
\end{align}
\end{subequations}
where $\Delta = h_{22}-h_{11}$, $\Omega = 2h_{12}$ and the sub-indices represent the row and column of matrix $h$. Eq.\eqref{5a} is derived from the commonly used approximate adiabatic criterion for non-oscillating Hamiltonians, which gives $V_0\ll8~E_r$. Typically, the optical lattice requires that the well depth $V_0$ has the same order with $E_r$. Thus, one cannot expect that the atoms move adiabatically ``forever". 
Eq.\eqref{5b} is used to limit the fast oscillating phase in $\Omega$, which can be solved as 
\begin{equation}\label{6}
	t \ll  \tau \triangleq \sqrt{\frac{8E_r}{V_0 k_L a}}
\end{equation}
where $\tau$ is the time bound for adiabatic evolution with the finite $V_0/E_r$. 

\subsection{Escape Mechanisms} \label{sec:2B}
\begin{figure}
	\centering
	\includegraphics[width=\linewidth]{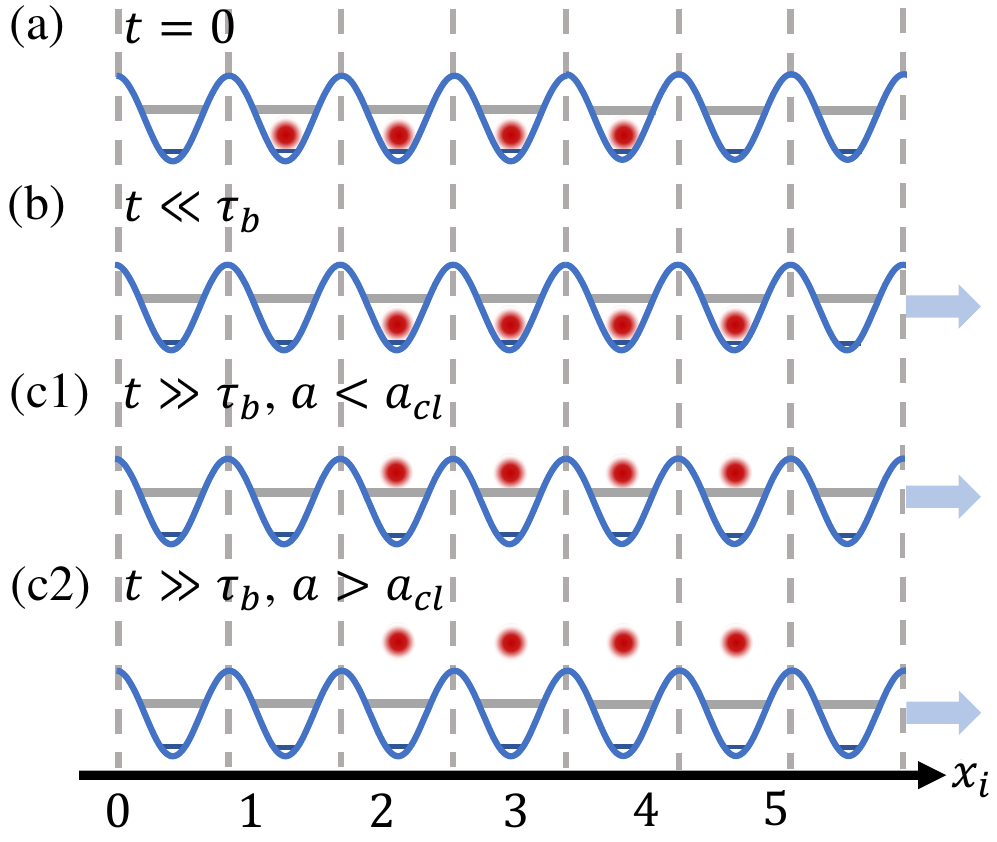}
	\caption{The schematic diagram of the relation between time bound and escape mechanisms. The blue solid lines represent the optical lattice. The gray lines represent different bands. The red shadows represent the atoms' distribution in space. The atoms are initially prepared in S-band, as shown in (a). The evolution of atoms bounded by inequality Eq.\eqref{9} is shown in (b). When the inequality Eq.\eqref{9} is not fulfilled, the atoms can populate either in higher bands (shown in (c1)) or in the free space (shown in (c2)). }
	\label{fig:1}
\end{figure}

In the above subsection, the time bound $\tau$ is derived as a part of adiabatic conditions and can be used to predict the time when adiabatic evolution breaks down. Suppose the atoms are initially prepared in the S-band. The instantaneous S-band evolves with a space translation for $at^2/2$ from the initial S-band. This space translation is exactly the displacement of optical lattice. Thus, roughly speaking, atoms that stay in the instantaneous S-band are also trapped in the lattice.

When the acceleration $a$ increases, the time bound $\tau$ exhibits a decrease characterized by a square root dependence on $a$. As $a$ goes to infinity, the Hamiltonian \eqref{1} can be reduced into the effective time-independent Hamiltonian:  
\begin{equation}\label{7}
	H_{\text{eff}} \triangleq \frac{1}{\Delta t}\int_{\tau}^{\tau + \Delta t} dt'\hat{H}(t')\xrightarrow{a\to\infty} \frac{\hat{P}^2}{2m} + \frac{V_0}{2}
\end{equation}
which is exactly the free particle Hamiltonian with constant potential $V_0/2$. Additionally, the wave functions of instantaneous bands collapse into the eigenstates of \eqref{7}, i.e., plane waves. Literally, the atoms cannot ``feel" the lattice. Before deriving the benchmark to scale the magnitude of acceleration $a$, we need to explain why the atoms can escape from the lattice.

From previous knowledge\cite{Bharucha1997, Wilkinson1997}, the decay of trapped atoms originates from two mechanisms. One is the quantum tunneling. Consider an atom is localized at a certain well. When the optical lattice is accelerated, the barriers will successively scatter this atom. In each scattering, the atom possesses a non-zero probability to tunnel through. And, the tunneled components are not accelerated by the lattice. When atomic relative velocity against the lattice increases, the tunneling probability increases polynomially. Thus, one could expect that, in the following scatterings, the atom becomes easier to tunnel through. As a result, the accumulation of tunneling contributes to a decay of trapped components.  

On the other hand, the atoms may fall down freely from the lattice due to its smooth shape when the acceleration $a$ is larger than some critical value. It can be illustrated by a classical model.
In the co-moving frame of reference, the potential that an atom feels is $V_0\sin^2\left(k_Lx\right)+max$. Thus, the acceleration of atom can be obtained by Newton's second law, namely, $\ddot{x} = -a+a_{cl}\sin\left(2k_Lx\right)$ where $a_{cl}\equiv {k_L V_0}/{m}$. For a classical atom, when $a<a_{cl}$, it is trapped in a certain well and cannot escape and is accelerated all the time; when $a>a_{cl}$, it escapes from this well. Therefore, $a_{cl}$ is the critical acceleration to identify the two escape mechanisms. When $a<a_{cl}$, the decay of trapped atoms is caused by quantum tunneling; when $a>a_{cl}$, the atoms classically fall down from the lattice. 

The critical acceleration $a_{cl}$ can be used as a benchmark to scale the magnitude of acceleration $a$. Suppose that $a\gg a_{cl}$, the atoms will quickly become untrapped, since then, the momenta distributions of atomic wave function become static. After that, the Hamiltonian Eq.\eqref{1} can be effectively replaced by a free particle Hamiltonian $H_{\operatorname{eff}}$. As a result, the adiabatic conditions in Eq.\eqref{5a} and \eqref{5b} are satisfied for $H_{\operatorname{eff}}$. It can also be understood by using the condition $a\gg a_{cl}$ in the expression of time bound $\tau$, i.e.,  
\begin{equation}\label{8}
	a\gg a_{cl}\;\Leftrightarrow \;\tau^2 \gg T_\mathrm{B}^2
\end{equation}
where the r.h.s of the arrow shows a square comparison of time bound and Bloch period.
Back to our original motivation, we expect that the time bound of adiabatic evolution faithfully reflects the time duration when atoms are trapped and accelerated. Thus, a large acceleration, i.e., $a\gg a_{cl}$ should be forbidden; then, the Hamiltonian Eq.\eqref{1} cannot be replaced by $H_{\operatorname{eff}}$. Therefore, the inequality Eq.\eqref{6} should be altered by adding a condition that $a\ll a_{cl}$, which gives
\begin{equation}\label{9}
	t \ll \tau,  ~\tau^2\ll T_\mathrm{B}^2.
\end{equation}

These inequalities generally limit the time bound less than one Bloch period. Within this time bound, the atoms are totally trapped and undergo adiabatic evolution. It should be stressed that Eq.\eqref{9} do not conflict with the fact that the atoms can perform many Bloch oscillations with a small enough $a/a_{cl}$ where the escape mechanisms are nearly negligible.

The above analysis can be visualized in \figurename\ref{1}. The atoms can be initially prepared in the S-band as shown in \figurename\ref{1}(a). Within the time bound Eq.\eqref{9}, the atoms are trapped and accelerated, as shown in \figurename\ref{1}(b). 
When the inequality Eq.\eqref{9} is not fulfilled, the atoms can populate either in higher bands (shown in
 \figurename \ref{fig:1}(c1) or in the free space (shown in  \figurename \ref{fig:1}(c2).

\section{Experimental Protocol}\label{sec:3}
\begin{figure}
	\centering
	\includegraphics[width=1\linewidth]{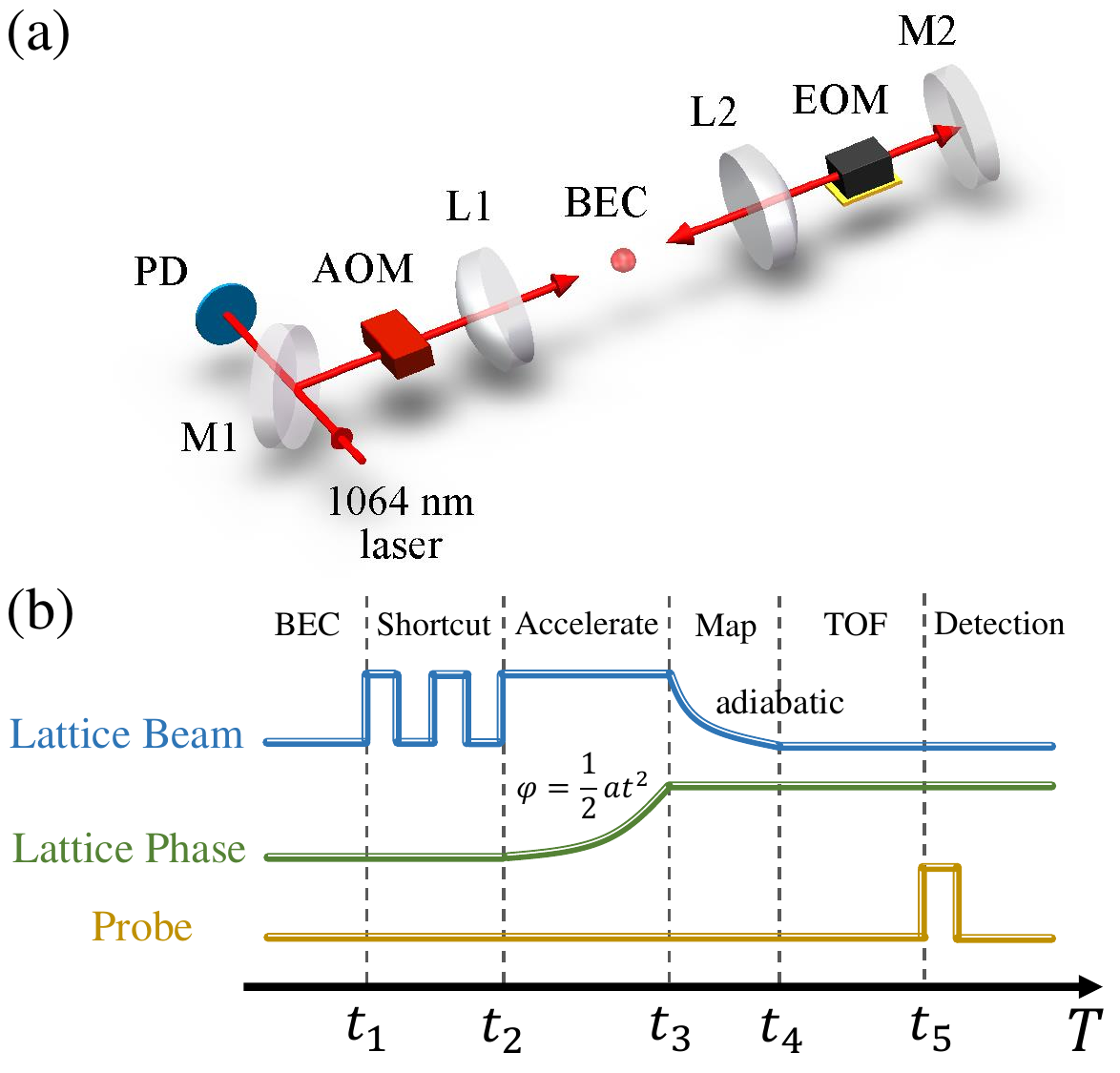}
	\caption{The experimental setups and sequence. (a) The experimental setups. The 1064nm laser beam is reflected by mirror to form the optical lattice, where the  AOM is to control the laser power and a minority of light is transmitted for feedback to stabilize the laser power. A phase modulation between the counter-propagating lattice beams is shaped through an EOM. The red solid lines with arrow indicate the optical path. (b) The time sequence diagram of instantaneous band measurements. The blue line represents the intensity of the lattice beam. It first performs the shortcut sequence from $ t_1$, then keep for the evolution time, and finally adiabatically turn off. The green line shows the phase of the lattice beam, and the phase is changed as $ a t^2/2$. The origin line expresses probe light that gives the absorption imaging after the time of flight.}
	\label{fig:2}
\end{figure}

To measure the time bound, we perform the experiments with the ultracold atoms in the 1D accelerating optical lattice.
We firstly prepare a dilute BEC of approximately $3\times 10^5$ ${}^{87}\text{Rb}$ atoms in a harmonic trap \cite{Jin2019,Yu2023,Hu2018}. Then we load the BEC into the S-band of the optical lattice using our proposed shortcut method \cite{Zhou2018}, where the optical lattice is created by a laser beam and its reflected beam with wavelength $1064\, ~\mathrm{nm}$ as shown in \figurename\ref{fig:2}(a). The shortcut method requires continuously turning on and off the lattice beams in order to modulate the initial state to the target state, where the whole process costs hundreds of $\rm \mu s$ with the fidelity near $1$. For example,  if the well depth is $6~E_r$ and the target state is S-band, the time sequence of operations is set as $42/6/54/24 ~\rm \mu s$ with fidelity $99.95 \%$.
In this process, the AOM (acousto-optic modulator) is used to realize the sharp variation of lattice depth in tens of microseconds with several pulses. Furthermore, we use the feedback control circuits to ensure that the laser intensity fluctuation is below $0.2\% @1s$. 

After the shortcut loading, we accelerate the optical lattice with a fixed acceleration $a$. It is achieved by using an EOM (electro-optic modulator, in \figurename \ref{fig:2} (a)) to modulate the phase of reflected lattice beam $\varphi$. The mode of modulation can be carefully chosen so that $\varphi = at^2/2$, i.e., the lattice is accelerated constantly, as shown in \figurename \ref{fig:2}(b). There, the phase noise of the accelerated optical lattice is attributed to the noise of the EOM control voltage within 50 mVpp, which gives rise to the acceleration uncertainty below $0.02\%$. 
Afterwards, the band distributions are observed using band mapping technique, where the intensity of lattice beams is adiabatically ramped down as an exponential curve in $1 ~\rm ms$. Finally, after the time of flight for $32 ~\rm ms$, the absorption imaging is taken to detect the atomic imaging.

In the experiment, to observe the obvious time bound, $V_0/E_r$ ranges from $10^0\sim10^1$ and $a$ ranges from $10^0\sim 10^4 ~\mathrm{m/s^2}$, as a result, the values of time bound $\tau$ vary from $10^0\sim 10^4 ~\rm \mu s$. Each distribution will be measured five times to obtain the statistical means and corresponding errors.

\begin{figure}
	\centering
	\includegraphics[width=0.9\linewidth]{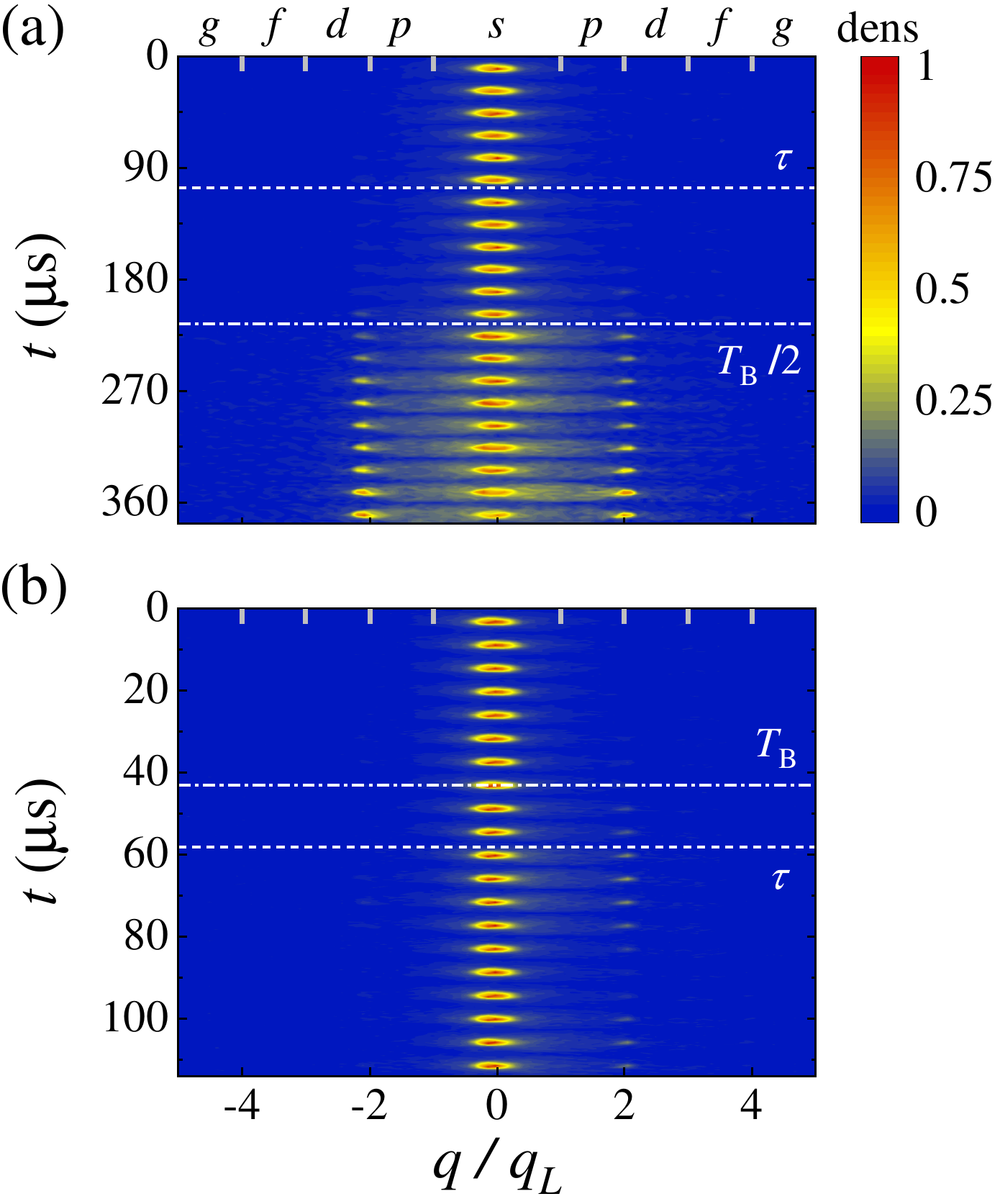}
	\caption{The typical images of atomic instantaneous band distributions. q is  defined as quasi momentum. The indices s, p, d, f, g represent the corresponding bands, respectively. Due to the small gap, the atoms that occupied in P-band and F-band are collected together in experiments. The white dashed lines represent the values of time bound $\tau$ and the white dot-dashed lines represent the values half of the Bloch period in (a) and Bloch period in (b). In (a), $V_0 = 6 ~E_r$, $a = 20 ~\mathrm{m/s^2}$ and $\tau/T_\mathrm{B}\approx 0.24$, $a/a_{cl} \approx 0.06$.  In (b), $V_0 = 2 ~E_r$, $a = 200 ~\mathrm{m/s^2}$ and $\tau/T_\mathrm{B}\approx 1.35$, $a/a_{cl} \approx 1.82$. }
	\label{fig:3}
\end{figure}

\section{Experimental results}\label{sec:4}
\subsection{The Instantaneous Band Distributions}\label{sec:4A}
To observe the adiabatic evolution of atoms in the accelerated optical lattice, the instantaneous band distributions of atoms are measured. The typical samples with selected parameters are shown in \figurename \ref{fig:3}. 
In \figurename \ref{fig:3} (a), the acceleration $a$ of the optical lattice is $20 ~\mathrm{m/s^2}$ and the lattice depth $V_0$ is $6~E_r$. Under these parameters, the time bound $\tau$ is about $0.24 T_\mathrm{B}$; the acceleration $a$ is about $0.06a_{cl}$. When the time $t$ is smaller than the time bound $\tau$, the atoms almost distribute in the instantaneous S-band, and the atoms are adiabatically accelerated by the optical lattice.
As the time $t$ increases to about half of Bloch period $T_\mathrm{B}/2$, an instantaneous inter-band transition takes place and the atoms tunnel to higher bands.

In \figurename \ref{fig:3}(b), the acceleration $a$ of the optical lattice is $200 ~\mathrm{m/s^2}$ and the lattice depth $V_0$ is $2~E_r$. Under these parameters, the time bound $\tau$ is about $1.35 T_\mathrm{B}$; the acceleration $a$ is about $1.82a_{cl}$. With the evolution time increasing, the atoms seem almost stay in the instantaneous S-band no matter if the time $t$ is smaller than time bound $\tau$. It is because the acceleration $a$ is so large that the instantaneous S-band function nearly collapses into plane waves after the time bound $\tau$ as illustrated in Section \ref{sec:2B}. 
 
\begin{figure}
	\centering
	\includegraphics[width=0.9\linewidth]{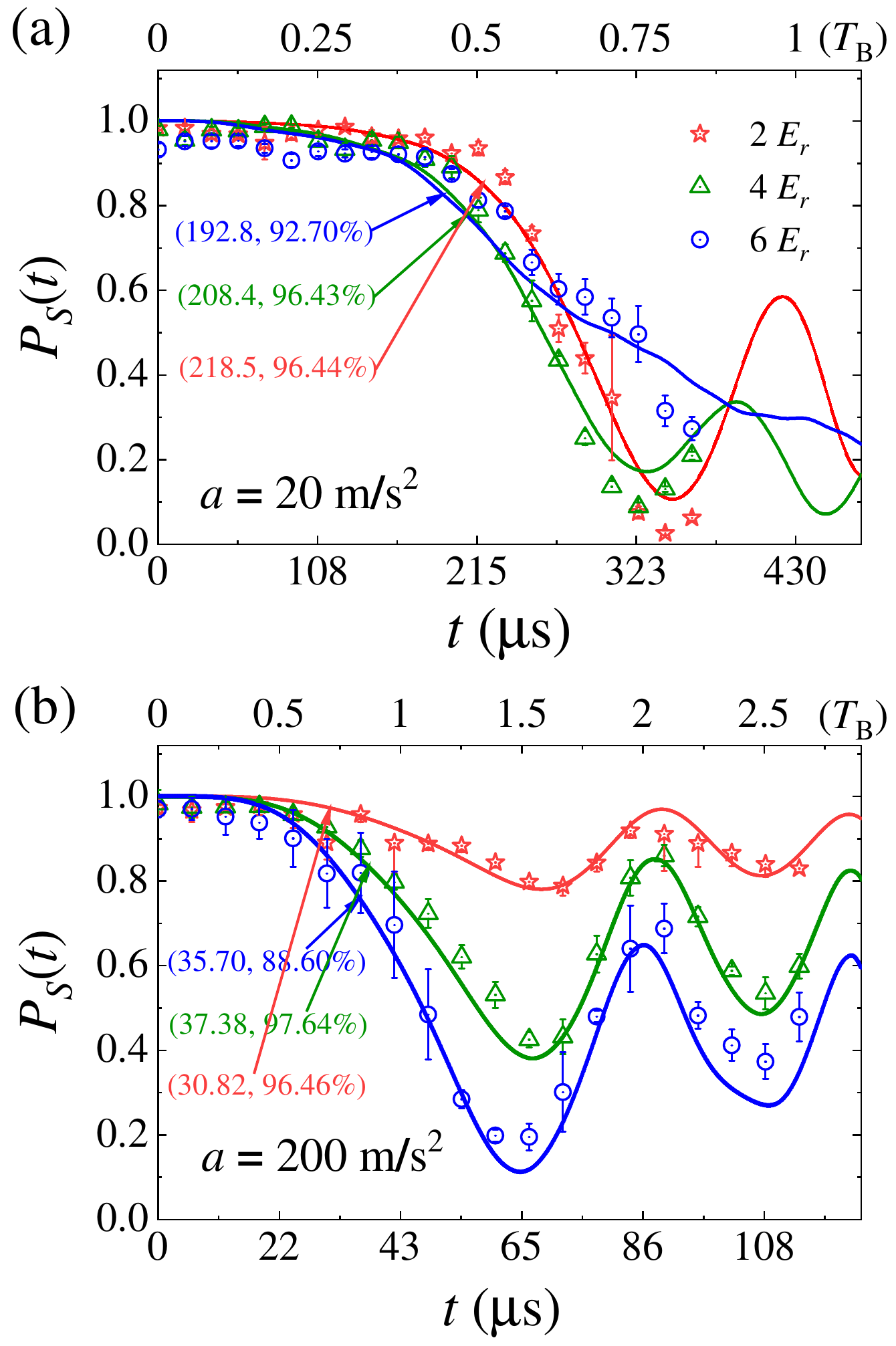}
	\caption{The instantaneous S-band distributions with different parameters. Red stars, blue triangles and green circles represent the experimental data with well depth $2 ~E_r$, $4 ~E_r$ and $6 ~E_r$, respectively. The solid lines represent the correspond numerical results from \eqref{1}. The arrows point out the drop point of $\ln P_S(t)$ of experimental results.  The accelerations used in (a) and (b) are $20 ~\mathrm{m/s^2}$ and $200 ~\mathrm{m/s^2}$, respectively.}
	\label{fig:4}
\end{figure}

To quantitative analyze atomic adiabatic evolution, we define the instantaneous band distributions $P_{\text{band}}(t)$ of atoms as 
\begin{equation}\label{10}
	P_{\text{band}}(t) = N_{\text{band}}(t)/N,
\end{equation}
where $N$ is the total number of atoms, and $N_{\text{band}}(t)$ is the atoms number in the target instantaneous band at time $t$.
The instantaneous S-band distributions $P_S(t)$ along time under different accelerations of lattice are shown in \figurename\ref{fig:4} (a) ($a = 20 ~\mathrm{m/s^2}$) and (b) ($a = 200 ~\mathrm{m/s^2}$). The solid lines are the results of numerical simulation on Hamiltonian Eq.\eqref{1}. The symbols are the results from experimental measurements.

In \figurename\ref{fig:4}(a), the atoms are driven by the optical lattice with acceleration $20\,\mathrm{m/s^2}$ and the well depths are selected as $2 ~E_r$ (red), $4 ~E_r$ (green) and $6 ~E_r$ (blue). The time bound $\tau$ defined in Eq.\eqref{6} are shown in the dashed vertical lines. When $t<\tau$, $P_S(t)$ are near to value $1$. The experimental $P_S(t)$ around the theoretical time bound are respectively $0.96$, $0.95$ and $0.93$ corresponding to well depth $2 ~E_r$, $4 ~E_r$ and $6 ~E_r$, which means the instantaneous inter-band transitions are depressed within $t<\tau$. 
Moreover, with an identical acceleration, when the well depth is larger, the time bound is smaller, and the adiabatic evolving time is shorter.
When $t>\tau$, $P_S(t)$ quickly drops to the individual converged values and oscillates around them.

In \figurename\ref{fig:4}(b), the acceleration of lattice is $200 ~\mathrm{m/s^2}$.
When $t<\tau$, the decay of $P_S(t)$ is also small, and the values of experimental data around individual $\tau$ are $0.84$, $0.80$ and $0.82$. 
Comparing to the situation in \figurename\ref{fig:4}(a), with the acceleration increasing, the time bound $\tau$ becomes shorter. 
Further, with an identical well depth, a larger acceleration of lattice gives a shorter adiabatic evolving time period. 
When $t>\tau$, $P_S(t)$ drops quickly, and the increase of acceleration also magnifies the oscillations of $P_S(t)$ with well depth $4 ~E_r$ (green) and $6 ~E_r$ (blue). 
But, the oscillations with well depth $2 ~E_r$ (red) is smaller than the situation with $a=20 ~\mathrm{m/s^2}$, and the red line shows an oscillating trend around a value that is approximate to $1$. For the green and blue lines, a finite proportion is left in the instantaneous S-band.

\subsection{The Measurement of Time Bound}\label{sec:4B}

\begin{figure}[H]
	\centering
	\includegraphics[width=0.9\linewidth]{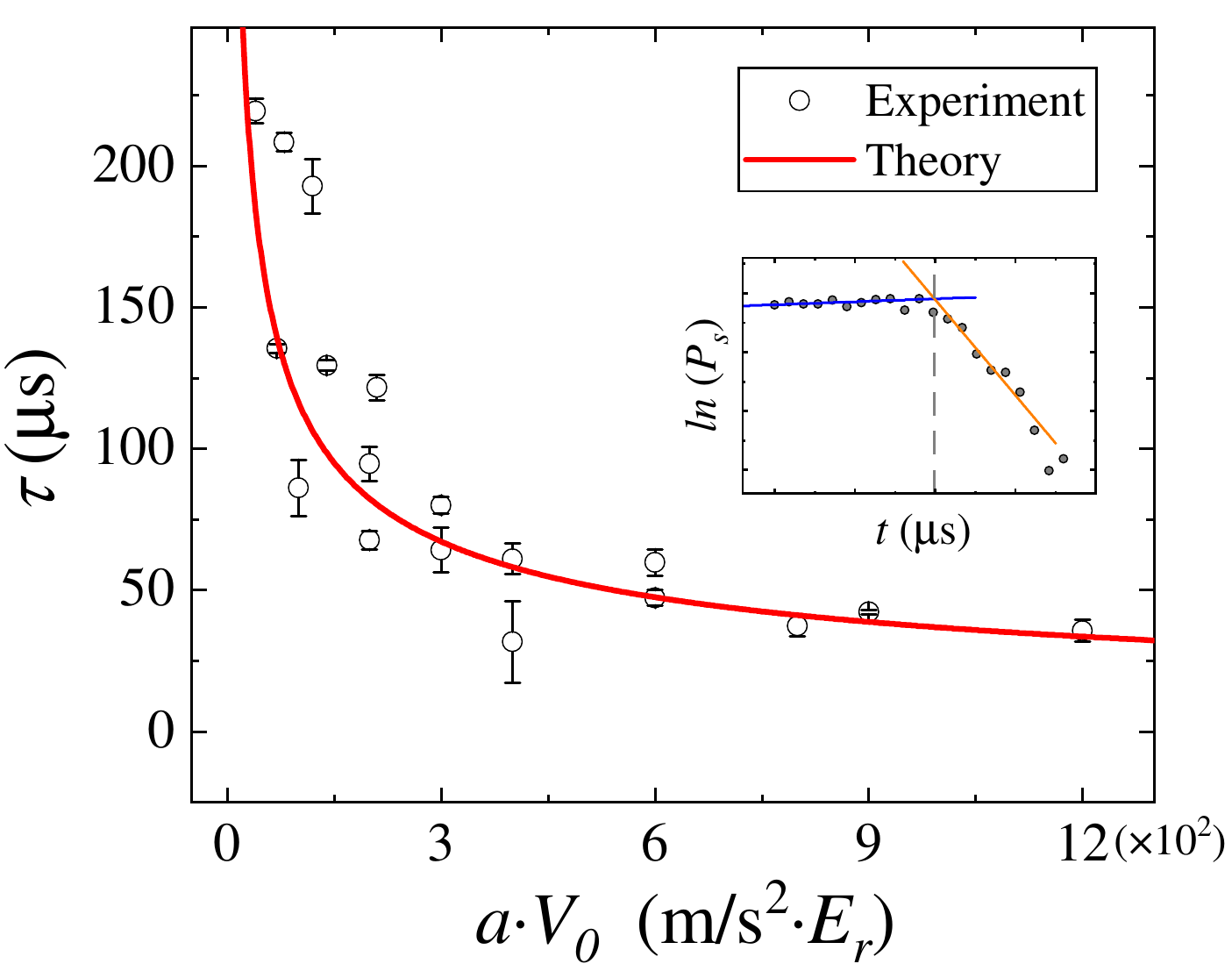}
	\caption{The relation between the time bound $\tau$ and the product $a \cdot V_0$. The red line is the theoretical result from Eq.\eqref{6} and the black circles are the experimental results. The error bar is the standard error of five repeated measurements. The time bound  is defined as the drop points of function $\ln P_S(t)$. The drop points are collected using the crossovers of two straight lines fitted by bilateral data points, as shown in the inserted figure.} 
	\label{fig:5}
\end{figure}

In order to verify the expression of time bound defined in Eq.\eqref{6}, we extract the time bound from the experimental data. We consider the time bound  as the time point where $P_S(t)$ rapidly decay. The insert of \figurename\ref{fig:5} shows our method to extract time bound $\tau$. Because the $P_S(t)$ suddenly decays around time bound, we use two lines to fit the $\ln(P_S)$ before and after the time bound respectively, as the blue and orange lines shown in the insert window of \figurename\ref{fig:5}. And, the intersection of two lines is extracted as the time bound.

\figurename\ref{fig:5} compares the experimental time bound and the theoretical prediction using different experimental parameters. The results show that the experimental data align reasonably well with the theoretical curve. Increasing the acceleration and lattice depth leads to a decrease in the time bound. If the product of the acceleration and lattice depth remains constant, the time bound remains the same regardless of variations in accelerations and lattice depths. However, it is important to note that there are still disparities between the experimental results and theoretical simulations, which can be attributed to several factors.
Firstly, in the experiment, the error in lattice depth calibration is within 5\%, while the error in acceleration calibration is approximately 2\% \cite{Yu2023}. These calibration errors contribute to the observed deviations.
Secondly, fluctuations in optical intensity and imperfections in the pulse waveform impact the experimental fidelity of the shortcut method, resulting in lower fidelity compared to the theoretical predictions.
Thirdly, the condensate of $^{87}\mathrm{Rb}$ exhibits particle interactions that introduce finite nonlinear effects to the system. To mitigate these effects, we select sufficiently large lattice potentials as an experimental parameter.

\section{The measurements of atomic velocity distributions}\label{sec}

\begin{figure}
	\centering
	\includegraphics[width=0.9\linewidth]{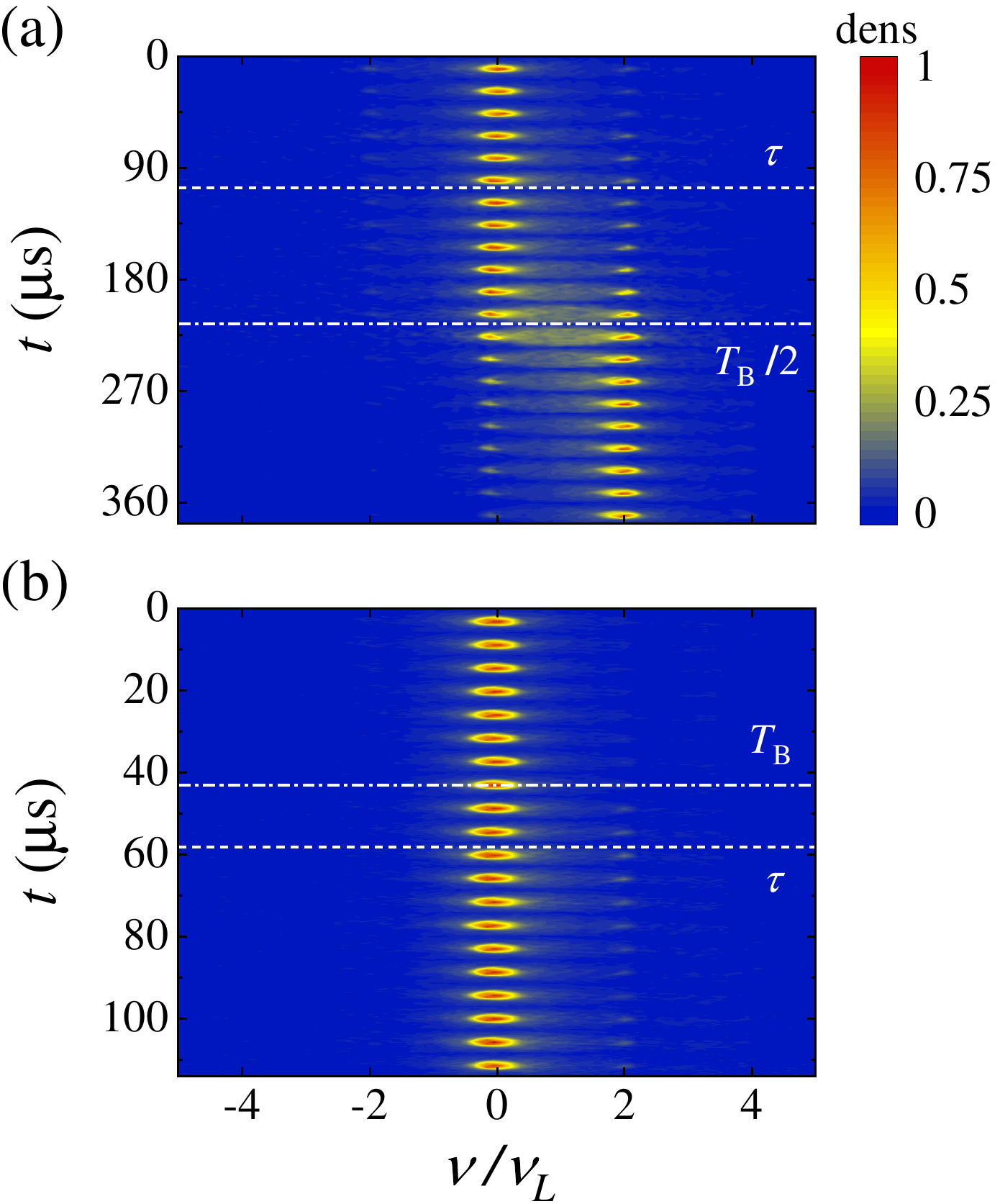}
	\caption{The typical images of the atomic velocity distributions. $v_L$ is defined as $\hbar k_L/m$. The white dashed lines represent the values of time bound $\tau$ and the white dot-dashed lines represent the values half of Bloch period in (a) and the Bloch period in (b). In (a), $V_0 = 6 ~E_r$, $a = 20 ~\rm m/s^2$ and $\tau/T_\mathrm{B}\approx 0.24$, $a/a_{cl} \approx 0.06$.  In (b), $V_0 = 2 ~E_r$, $a = 200 ~\rm m/s^2$ and $\tau/T_\mathrm{B}\approx 1.35$, $a/a_{cl} \approx 1.82$.}
	\label{fig:6}
\end{figure}

We also perform the measurements of atomic velocity distributions at different time in order to study its relation with time bound $\tau$. The typical images are shown in \figurename \ref{fig:6}. The atoms are initially prepared in the S-band, where most of atoms distribute at a zero-velocity state, and a small part of atoms distribute at the $\pm 2v_L$ states. In \figurename \ref{fig:6}(a), where $a\ll a_{cl}$, the atoms are accelerated, and a velocity eigenvalue transfer (with amplitude $2v_L$) can be observed around $T_\mathrm{B}/2$ which is regarded as one character of Bloch oscillation.

In \figurename \ref{fig:6}(b), where $a> a_{cl}$, the velocity distribution of atoms preserves the same for more than one Bloch period. It shows that the atoms are not accelerated by the lattice because they have almost escaped from the lattice, which is illustrated in Section \ref{sec:2B}.

\begin{figure}
	\centering
	\includegraphics[width=0.8\linewidth]{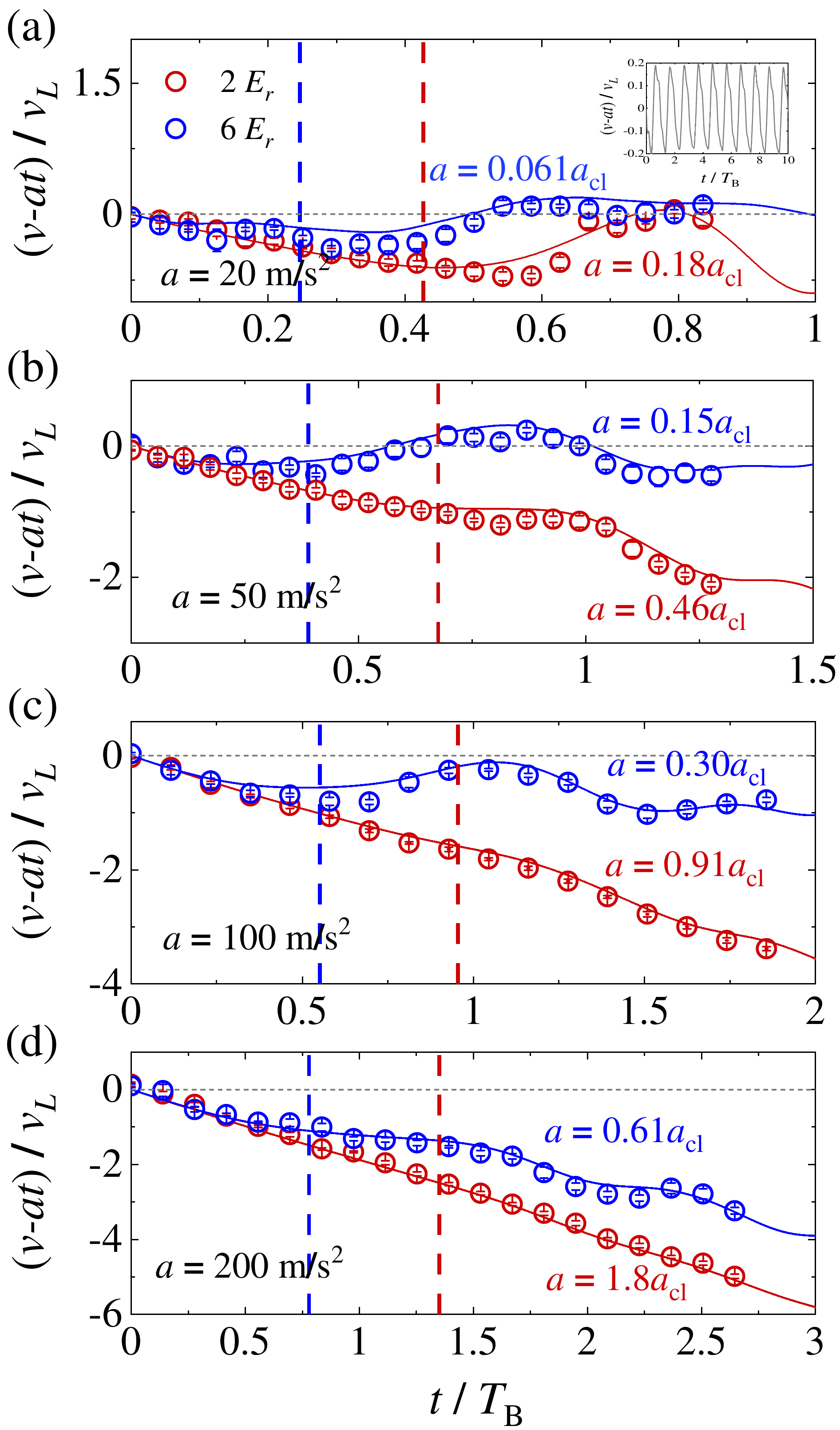}
	\caption{The atomic relative velocity against the lattice's velocity. The blue lines and the red lines are respectively the experimental data for well depth $2 ~E_r$ and $6 ~E_r$. The gray short dash lines represent the velocity reference of the accelerated lattice. \figurename (a) to (d) are set with different accelerations, i.e., $a = 20$, $50$, $100$, $200 ~\mathrm{m/s^2}$. Inset in (a): the numerical simulation of atomic relative velocity for $10$ Bloch periods with $a/a_{cl} = 0.061$.}		
	\label{fig:7}
\end{figure} 
 
The atomic relative velocity against the lattice's velocity with different parameters are shown in \figurename\ref{fig:7}. 
In \figurename \ref{fig:7}(a), the blue circles are the relative velocity with {$a=0.061a_{cl}$}, and it oscillates up and down around zero (denoted by the gray short dash line) for more than one Bloch period. Besides, its crossover with zero is at the half of Bloch period. 
It is a sign that the atoms are performing Bloch oscillations. In the inset of \figurename \ref{fig:7}(a), we calculate numerically further for longer time, which shows the Bloch oscillations last over $10$ Bloch periods with less distortion. It means that the time bound does not prevent the Bloch oscillation over many periods under such small $a/a_{cl}$.In the experiment, due to limited by the maximum control voltage of EOM, we can't measure any more Bloch periods.
The red circles are the relative velocity with { $a=0.18a_{cl}$}, and they also stay very close to zero. However, because the crossover is not at the half of Bloch period, it is not a standard Bloch oscillation and the atoms are slightly left behind the lattice. 
In \figurename\ref{fig:7}.(b) and (c), the acceleration $a$ is near $a_{cl}$. The blue circles show deformed oscillations with the crossovers drifting because some of the components have escaped from the lattice. The red circles fall far behind the lattice and show a trend to keep a finite velocity. It is a signal that a large component of atoms is not accelerated and they move like free particles with a static velocity distribution. In this way, the expectation of atoms' velocity becomes static since the escape. 
In \figurename\ref{fig:7}.(d), $a$ is increased to $200 ~\mathrm{m/s^2}$, which is larger than the classical critical velocity $a_{cl}$ of the red circles. The clustering of red circles follows the velocity of lattice without oscillations, which illustrates the absence of atom acceleration.

\section{conclusions}\label{sec:7}
The present study shows that the atomic adiabatic evolution in the accelerated optical lattice is limited by a time bound $\tau$. This time bound is obtained through a two-momenta state approximation and is expressed in Eq. \eqref{6}. The theoretical framework is verified through measurements of the instantaneous band distributions at different times $P_S(t)$ with varying lattice's depths and accelerations, as shown in \figurename\ref{fig:4}. The drop points of $P_S(t)$ are fitted and collected to compare with the theoretical prediction. And it shows that they agree well in \figurename\ref{fig:5}. To further ensure that the atoms keep trapped within the lattice during the adiabatic evolution, a limit on the acceleration of the optical lattice is necessary. Specifically, it is recommended that the acceleration be much smaller than a characteristic acceleration $a_{cl}$, as indicated in Eq. \eqref{8} and \eqref{9}. The verification of this condition is demonstrated through the observation of atomic velocity distributions at various times. 

In many experimental cases, the accelerated optical lattice is used to accelerate atoms. However, during the acceleration process, most atoms become untrapped and fall freely due to escape mechanisms. Additionally, the proportion of untrapped atoms increases as the final atomic velocity becomes larger. This highlights the challenge of maintaining a high population of trapped atoms during the acceleration process. The derived time bound helps to optimize the experimental parameters to strike a balance between achieving a desirable final atomic velocity and preserving a sufficient population of trapped atoms for subsequent experiments.

It should be noted that the characteristic acceleration $a_{cl}$ is derived from our understanding of escape mechanisms rather than detailed calculations. So, its validity and limitations need further discussion in the future studies.

\section{acknowledgments}\label{sec:8}
The authors thank professor Biao Wu for helpful comments.
This work was supported by the National Key Research and Development
Program of China (Grants No. 2021YFA0718300 and
No. 2021YFA1400900),  the National Natural
Science Foundation of China (Grants No. 11934002 and
No. 11920101004), the Science and Technology Major
Project of Shanxi (Grant No. 202101030201022), the National Key R\&D
Program of China (Grant No. 2018YFA0305602), the Space Application System of China Manned Space Progra.

Guoling Yin and Lingchii Kong contributed equally to this work.

\end{document}